# Prediciendo el Generador Cuadrático

**Domingo Gómez-Pérez, Jaime Gutiérrez, Álvar Ibeas y David Sevilla**

Facultad de Ciencias
Universidad de Cantabria
Santander E-39071, España
jaime.gutierrez@unican.es

**Abstract.** Sean $p$ un primo, $a$ y $c$ enteros módulo $p$ tales que $a \neq 0 \bmod p$. El generador cuadrático es una sucesión ($u_n$) de números pseudoaleatorios definidos por la relación $u_{n+1} \equiv_p a u_n^2 + c$. En este trabajo demostramos que si conocemos un número suficientemente grande de los bits más significativos para dos valores consecutivos $u_n, u_{n+1}$, entonces podemos descubrir en tiempo polinomial la semilla $u_0$, excepto para un conjunto pequeño de valores excepcionales.

## 1. Introducción

En el corazón de gran parte de los sistemas criptográficos usados actualmente está la generación de números secretos o aleatorios, que no puedan ser adivinados para la seguridad del criptosistema. La motivación de la generación de números pseudoaleatorios es que en algunos casos necesitamos manipular más bits aleatorios de los que pueden proporcionarnos nuestras fuentes físicas de entropía. En estos casos se recurre a los PRNG, generadores de números pseudoaleatorios, que constituyen, en definitiva, una forma de expandir unos pocos bits realmente aleatorios.

Un PRNG es una transformación que toma cierta cantidad aleatoria, llamada semilla, y genera una secuencia de bits que pueden usarse como si fuesen números "casi" aleatorios. Un campo muy activo es, precisamente, estudiar y analizar cuánto de "aleatoriedad" posee la sucesión obtenida.

Un bien conocido y estudiado PRNG, pero muy inseguro criptográficamente, es el generador congruente lineal, definido por un polinomio lineal. Nosotros consideramos un generador definido por un polinomio cuadrático.

Para un primo $p$, denotamos por $F_p$ el cuerpo de $p$ elementos y, como siempre, asumiremos que está representado por el conjunto $\{0,..., p-1\}$. Algunas veces trataremos los elementos de $F_p$ como enteros en ese rango.

Fijados $a \in F_p^*$ y $c \in F_p$, consideramos el polinomio $f(X) = aX^2 + c \in F_p[X]$. Definimos el generador congruente cuadrático $(u_n)$ de elementos de $F_p$ por la relación recurrente:

$$u_{n+1} \equiv f(u_n) \mod p \qquad n = 0,1,... \qquad (1)$$

donde $u_0$ es el valor inicial o semilla. Las constantes $a$ y $c$ son denominadas el multiplicador y el desplazamiento, respectivamente, del generador.

Este trabajo analiza algunas propiedades criptográficas de este generador, que incluye el caso de multiplicador uno, que se corresponde con el célebre generador de Pollard.

En las aplicaciones a la criptografía, la semilla $u_0$ y las constantes $a$ y $c$ se suponen parte de la clave secreta. Se quiere usar la salida del generador como un cifrado de flujo. Por supuesto, si varios valores consecutivos son revelados, entonces es muy fácil descubrir $u_0$, $a$ y $c$. De esta forma, solamente se envían los bits más significativos de cada $u_n$ con la esperanza de que sea difícil predecir la sucesión. En el reciente trabajo [2] se prueba que no muchos bits pueden darse en cada etapa: desafortunadamente, el generador cuadrático es predecible si se revelan un número suficientemente grande de los bits más significativos de elementos consecutivos, tan grande como un número menor de claves que son excluidas. Sin embargo, la mayor parte de los resultados en [2], requieren excluir una conjunto pequeño de pares $(a,c)$. Si este conjunto pequeño no es excluido, el algoritmo para encontrar la información secreta puede fallar. En principio, se puede esperar que por una elección deliberada del par $(a,c)$ en este conjunto excluido pueda generar sucesiones criptográficamente más seguras. El objetivo de este artículo es demostrar que esta estrategia no es exitosa. De hecho, nosotros introducimos algunas modificaciones y añadidos al método de [2] de tal forma que permite atacar los generadores, independientemente de donde se elijan los valores $a$ y $c$. Mostraremos este estudio en el caso en que $a$ y $c$ son conocidos. El supuesto de que $a$ y $c$ son públicos reduce la relevancia del problema en criptografía. Pero creemos que la mejora del resultado es interesante en sí misma. Además, también creemos que nuestra técnica puede ser extendida cuando $a$ y $c$ son ambos secretos.

Asumimos que la sucesión $(u_n)$ no es conocida, pero para algún $n$, son dadas dos aproximaciones $w_0$ y $w_1$ de dos valores consecutivos $u_n$ y $u_{n+1}$. Demostraremos que podemos descubrir en tiempo polinomial los valores $u_n$ y $u_{n+1}$ si las aproximaciones son suficientemente buenas y si un conjunto pequeño de valores de $u_0$ es excluido. (El trabajo en [2] excluye, además del conjunto para $u_0$, un conjunto pequeño de valores de $(a,c)$, en este sentido nuestro resultado es más fuerte.

Problemas similares han sido introducidos por Knuth [10] para el generador congruente lineal

$$x_{n+1} \equiv ax_n + c \mod p \qquad n = 0,1,...$$

y después considerados en [3,4,7,11], ver también los trabajos [5,12].

En alguno de estos trabajos se han considerado generadores no lineales, pero sólo cuando todos los términos se han dado completamente (ver [12]). También se ha estudiado el problema para el generador congruente inverso (ver [17]).

$$x_{n+1} \equiv a x_n^{-1} + c \quad \mod p \quad n = 0,1,...$$

En todo el trabajo, las palabras tiempo polinomial significan polinomial en $\log p$. Los resultados involucran otro parámetro $\Delta$, el cual mide cuanto de aproximados son los valores $w_j$ en términos de $u_{n+j}$. Este parámetro varía independientemente de $p$ sujeto a la desigualdad $\Delta < p$, y no aparece en la estimación de la complejidad del algoritmo presentado.

Debemos señalar que el algoritmo presentado es riguroso y determinista (ver [2] para una discusión rigurosa y heurística de este tipo de algoritmos).

Comienza el artículo con un breve repaso de los resultados básicos sobre retículas en la Sección 2.1 y, polinomios en la Sección 2.2. En la Sección 3.1 formulamos el resultado principal y un esquema de demostración, la cual aparece en la Sección 3.2. Finalmente, la Sección 4 está dedicada a presentar algunas comentarios concluyentes y a presentar problemas abiertos.

## 2. Retículas y Polinomios

### 2.1. Sumario sobre Retículas

En esta subsección recogemos varios resultados conocidos sobre retículas, que forman los antecedentes para nuestros algoritmos.

Revisamos varios resultados y definiciones relacionados sobre retículas (lattices), que pueden encontrarse en [6]. Para más detalles y referencias más recientes, recomendamos consultar [1,7,8,14-16].

Sea $b_1,...b_s$ un conjunto de vectores linealmente independientes en $R^r$. El conjunto

$$L = \{c_1 b_1 + ... + c_s b_s \,/\, c_1,...,c_s \in Z\}$$

se llama retícula $s$-dimensional con base $\{b_1,...,b_s\}$. Si $s = r$, la retícula $L$ es de rango máximo.

A cada retícula $L$ uno puede asociar de manera natural su volumen

$$vol(L) = \sqrt{\det(\langle b_i, b_j \rangle)_{i,j=1}^{s}}$$

donde $\langle a,b \rangle$ denota el producto escalar, que no depende de la elección de la base $\{b_1,...,b_s\}$.

Dado un vector $u$, sea $\|u\|$ su norma euclídea. El famoso Teorema de Minkowski, (ver Teorema 5.3.6 en la Sección 5.3 de [6]), proporciona una cota superior

$$\min\{\|z\|: z \in L \setminus \{0\}\} \leq s^{1/2} vol(L)^{1/s} \qquad (2)$$

de un vector no nulo con norma mínima, en una retícula $s$-dimensional $L$ en términos de su volumen. De hecho, $s^{1/2}$ puede ser substituido por la constante de Hermite $\gamma_s^{1/2}$, para la que tenemos

$$\frac{1}{2\pi e}s + o(s) \leq \gamma_s \leq \frac{1.744}{2\pi e}s + o(s), \qquad s \to \infty$$

La cota de Minkowski (2) motiva una pregunta natural: ¿cómo encontrar un vector no nulo y con norma mínima en una retícula? Diremos que es un vector corto de la retícula. El célebre algoritmo LLL de Lenstra, Lenstra y Lovász [13] proporciona una solución deseable en la práctica, y se sabe que el problema es resoluble en tiempo polinomial determinista (polinomial en el tamaño-bit de la base de $L$) suponiendo que la dimensión de $L$ esté fijada (ver Kannan [9, Section 3], por ejemplo). Las retículas en este artículo tienen dimensión fija (nótese que se conocen varios indicios de que el problema del vector más corto es NP-completo cuando la dimensión crece).

De hecho, en este artículo sólo consideramos retículas muy especiales. Concretamente, retículas que consisten en soluciones enteras $x = (x_0,...,x_{s-1}) \in Z^s$ del sistema de congruencias

$$\sum_{i=0}^{s-1} a_{ij} x_i \equiv 0 \bmod q_j, \qquad j=1,...,m$$

módulo algunos enteros $q_1,...,q_m$. Típicamente (aunque no siempre) el volumen de una de estas retículas es el producto $Q = q_1 \cdots q_m$. Aún más, todos los algoritmos mencionados anteriormente, cuando son aplicados a una de estas retículas, se convierten en polinomiales en $\log Q$.

### 2.2. Ceros de polinomios

Nuestra segunda herramienta básica es esencialmente el teoremade Lagrange que afirma que un polinomio no nulo de grado $N$ sobre cualquier cuerpo no tiene más de $N$ ceros en ese cuerpo.

Los polinomios que consideramos pertenecen a una cierta familia de funciones parametrizada por vectores con norma pequeña en una cierta retícula, por tanto el tamaño de la familia se puede mantener bajo control. Los ceros de esos polinomios lineales se corresponden a valores iniciales potencialmente "malos" del generador cuadrático (1). Por tanto, si todos los polinomios lineales en esta familia no son idénticamente cero módulo $p$ entonces tenemos una cota superior para el número de valores iniciales "malos". Por tanto, una parte crucial de nuestra aproximación es estudiar la posible anulación de polinomios lineales en la familia anterior y mostrar que esto puede ocurrir sólo para muy pocos valores de los coeficientes del generador (1).

## 3. Prediciendo el Generador Cuadrático con Multiplicador y Desplazamiento Públicos

### 3.1. Formulación del Resultado Principal y Plan de Demostración

Supongamos que el multiplicador $a$ y el desplazamiento $c$ del generador cuadrático son conocidos. Mostraremos que podemos recuperar $u_0$ para todos sus posibles valores, salvo un conjunto de $O(\Delta^4)$ elementos, cuando se dan dos $\Delta$-aproximaciones a dos valores consecutivos $u_n, u_{n+1}$ producidos por el generador cuadrático. Para simplificar la notación, supondremos que $n = 0$ a partir de este punto.

**Teorema.** Sean $p$ un número primo, $a \in F_p^*, c \in F_p$ y $\Delta \in Z$ con $1 \leq \Delta < p$. Entonces existe un conjunto $U(\Delta; a, c) \subseteq F_p$ con cardinal $\#U(\Delta; a, c) = O(\Delta^4)$, con la siguiente propiedad: siempre que $u_0 \notin U(\Delta; A, c)$, dadas aproximaciones $w_j$ tales que

$$|w_j - u_j| \leq \Delta, \qquad j = 0,1$$

a dos valores consecutivos $u_0, u_1$ producidos por el generador cuadrático (1), se puede calcular $u_0$ en tiempo polinomial determinista.

A continuación mostramos el esquema del algoritmo correspondiente a la demostración de este Teorema. El algoritmo está dividido en seis etapas:

**Etapa 1**: Construimos una cierta retícula $L$ (ver (4) más abajo) de dimensión cuatro que depende de $w_0, w_1$ y $a, c$. Se muestra que un cierto vector $e$, relacionado con información oculta que buscamos sobre $u_0, u_1$, es un vector "muy" corto en esta retícula. Calculamos un vector no nulo corto $f = (f_0, \ldots, f_3)$ en $L$; ver [9] para el algoritmo correspondiente.

**Etapa 2**: Mostramos que $f$ proporciona información valiosa sobre $e$ para todos los posibles valores iniciales $u_0$ excepto aquellos en un cierto conjunto excepcional $V(\Delta; a, c) \subseteq F_p$ de cardinal $\#V(\Delta; a, c) = O(\Delta^4)$, que se define como el conjunto de ceros de una cierta familia paramétrica de polinomios lineales.

**Etapa 3**: Mostramos que si $f_0 \neq 0$ entonces recuperar $e$ (y por tanto la información oculta $u_0$) es inmediato. Por tanto en ese caso el algoritmo termina en esta etapa.

**Etapa 4**: Mostramos que si $f_0 = 0$ entonces el vector $f$ nos permite calcular enteros $r$ y $s$ tales que $b \equiv r/s \mod p$ (de hecho estos enteros se pueden calcular independientemente con el algoritmo de fracciones continuas). Usamos esta información, junto con los enteros $w_0, w_1$, para calcular una segunda retícula $L'$ (ver (5)) de dimensión tres. De nuevo hay un vector "muy" corto $e'$ en $L'$ que está estrechamente relacionado con la información secreta $u_0$.

**Etapa 5**: Mostramos que todos los vectores cortos en $L'$ son paralelos a $e'$ para todos los valores posibles de $u_0$ salvo los de otro conjunto excepcional $V'(\Delta; a, c) \subseteq F_p$ de cardinal $V'(\Delta; a, c) = O(\Delta^4)$ (que también se define como el conjunto de ceros de una cierta familia paramétrica de polinomios lineales).

**Etapa 6**: Encontramos un vector no nulo corto $f'$ en $L'$ y mostramos que si $u_0 \notin U(\Delta; a, c)$, donde $U(\Delta; a, c) = V(\Delta; a, c) \cup V'(\Delta; a, c)$ es trivial recuperar $e'$ (y por tanto encontrar la información secreta) a partir de $f$ y $f'$.

### 3.2. Demostración del resultado principal

El teorema es trivial cuando $\Delta^4 \geq p$, luego supondremos que $\Delta^4 < p$. Supondremos que $u_0 \notin U(\Delta; a, c)$, conjunto que definiremos gradualmente a lo largo de la demostración.

*Etapa 1: Construcción de la retícula $L$.*

Comenzamos definiendo $L$, y mostrando cómo un vector corto "normalmente" permite recuperar $u_0$.

Sean $w_0, w_1$ las dos aproximaciones de las que partimos. Se verifica entonces que hay dos enteros $\varepsilon_0, \varepsilon_1$ con:
$$u_i = w_i + \varepsilon_i, |\varepsilon_i| \leq \Delta \qquad i = 0,1$$

Tenemos que:
$$u_1 \equiv_p f(u_0) \Rightarrow w_1 + \varepsilon_1 \equiv_p a(w_0 + \varepsilon_0)^2 + c \Rightarrow$$
$$\underbrace{aw_0^2 + c - w_1} + \underbrace{2aw_0}\varepsilon_0 + \underbrace{(-1)}\varepsilon_1 + \underbrace{a}\varepsilon_0^2 \equiv_p 0 \qquad (3)$$

Por lo tanto, si $L$ es la retícula definida por el siguiente sistema de congruencias:

$$L : \begin{cases} (aw_0^2 + c - w_1)\Delta^{-2}x_0 + 2aw_0\Delta^{-1}x_1 - \Delta^{-1}x_2 + ax_3 \equiv 0 \mod p \\ x_0 \equiv 0 \mod \Delta^2 \\ x_1, x_2 \equiv 0 \mod \Delta \end{cases} \qquad (4)$$

se tiene que el siguiente vector de $Z^4$ está en $L$:

$$e = (\Delta^2, \Delta\varepsilon_0, \Delta\varepsilon_1, \varepsilon_0^2)$$

Este vector satisface $\|e\| \leq \sqrt{4\Delta^4} = 2\Delta^2$. Sea $f = (\Delta^2 f_0, \Delta f_1, \Delta f_2, f_3)$ un vector corto de la retícula, entonces $\|f\| \leq \|e\| \leq 2\Delta^2$ y, por tanto $|f_0| \leq 2, |f_1|, |f_2| \leq 2\Delta, |f_3| \leq 2\Delta^2$.

Obsérvese que el vector $f$ puede ser computado en tiempo polinomial, a partir de los datos de entrada.

*Etapa 2: Definición del primer conjunto excepcional $V(\Delta; a, c)$.*

Consideramos un tercer vector de $L$:

$$d = f - f_0 e = (0, \Delta d_1, \Delta d_2, d_3)$$

Se puede esperar que $d$ sea el vector nulo, es decir, los vectores $e$ y $f$ sean paralelos. Desafortunadamente, esto no es cierto en general, demostraremos que es cierto si $d_1 \neq 0$ y $v_0$ no pertenece al conjunto $V(\Delta; a, c)$, que definiremos más adelante.

Por la definición de $L$,

$$2aw_0 d_1 - d_2 + a d_3 \equiv 0 \bmod p$$

Substituyendo $w_0 = u_0 - \varepsilon_0$ en esta ecuación, obtenemos:

$$M(u_0) = 2a d_1 u_0 \equiv E \bmod p$$

donde $E \equiv a(2d_1 \varepsilon_0 - d_3) + d_2 \bmod p$. Si el polinomio $M(u_0)$ es no constante, es decir, $d_1 \neq 0 \bmod p$, entonces, fijado $E$ hay un único $u_0$ verificando esa ecuación.

Definimos ahora $V(\Delta; a, c)$ como el conjunto de los elementos $v \in F_p$ para los que existe alguna combinación $(d_1, d_2, d_3, \varepsilon_0) \in Z^* \times Z^3$ que satisface

$$|d_1|, |d_2| \leq 4\Delta, |d_3| \leq 4\Delta^2, |\varepsilon_0| \leq \Delta$$
$$2a d_1 v \equiv E \bmod p$$

Como la cantidad de valores que $E$ puede tomar está en $O(\Delta^3)$, se sigue que $\#V(\Delta; a, c) = O(\Delta; a, c)$.

Ahora, si $u_0 \notin V_1(\Delta; a, c)$, entonces debe ser $d_1 = 0 \Rightarrow f_0 \varepsilon_0 - f_1 = 0$. Además, tenemos que

$$-d_2 + a d_3 \equiv 0 \bmod p$$

*Etapa 3: Predicción del generador cuando $f_0 \not\equiv 0 \bmod p$.*

Podemos calcular a partir de los datos conocidos

$$\varepsilon_0 = f_1 f_0^{-1}$$

Por tanto, conocemos $u_0 = w_0 + \varepsilon_0$ y, aplicando el generador cuadrático, $u_1$.

*Etapa 4: Construcción de la retícula $L'$.*

Sabemos que $|f_0| \le 2$. Por lo tanto, debe ser $f_0 = 0$. Así,
$$d = f = (0, 0, \Delta f_2, f_3)$$
y se verifica que $f_2 \equiv a f_3 \bmod p$.

Computamos entonces los enteros coprimos siguientes:

$$r := \frac{f_2}{mcd(f_2, f_3)}, \qquad s := \frac{f_3}{mcd(f_2, f_3)}$$

que verifican $r \equiv as \bmod p, |r| \le 2\Delta, |s| \le 2\Delta^2$.

De la ecuación (3) deducimos que

$$\underbrace{rw_0^2 - sw_1 + sc}_{} + \underbrace{2rw_0 \varepsilon_0}_{} - s\varepsilon_1 + r\varepsilon_0^2 \equiv 0 \bmod p$$

Con los datos ahora conocidos, consideramos la siguiente retícula:

$$L': \begin{cases} (rw_0^2 + sc - sw_1)\Delta^{-3} x_0 + 2rw_0 \Delta^{-2} x_1 + x_2 \equiv 0 \bmod p \\ x_0 \equiv 0 \bmod \Delta^3 \\ x_1 \equiv 0 \bmod \Delta^2 \end{cases} \qquad (5)$$

Sabemos que contiene el vector

$$e' = (\Delta^3, \Delta^2 \varepsilon_0, r\varepsilon_0^2 - s\varepsilon_1)$$

Además, $\|e'\| \le \sqrt{\Delta^6 + \Delta^6 + 16\Delta^6} = 3\sqrt{2}\Delta^3$.

*Etapa 5: Definición del segundo conjunto excepcional $V'(\Delta; a, c)$.*

Demostraremos que todos los vectores cortos de $L'$ son paralelos a $e'$, a menos que $u_0$ pertenezca a un conjunto $V'(\Delta; a, c)$ que detalleremos.

Supongamos lo contrario, luego existe un vector $f' \in L'$ no paralelo al vector $e'$ con $\|f'\| \leq \|e'\| < 3\sqrt{2}\Delta^3$, que tendrá la forma
$$f' = (\Delta^3 f_0', \Delta^2 f_1', f_2')$$

Podemos acotar sus componentes de manera paralela a la etapa anterior:
$$|f_0'| < 3\sqrt{2} \Rightarrow |f_0'| \leq 4, |f_1'| < 3\sqrt{2}\Delta, |f_2'| < 3\sqrt{2}\Delta^2$$

Consideramos un tercer vector de esta nueva retícula:
$$d' := f' - f_0' e' = (0, \Delta^2 d_1', d_2')$$

Sus componentes han de satisfacer
$$|d_1'| < 3\sqrt{2}\Delta + 4\Delta \Rightarrow |d_1'| < 9\Delta, |d_2'| < 3\sqrt{2}\Delta^3 + 16\Delta^3 \Rightarrow |d_2'| < 21\Delta^3$$

Por otro lado, por estar en la retícula,
$$2rw_0 d_1' + d_2' \equiv 0 \bmod p \tag{6}$$

Utilizando las cotas, tenemos que si $d_1' \equiv 0 \bmod p$, entonces $d_1' = d_2' = 0$ y los vectores $e'$ y $f'$ son paralelos. Esto contradice la elección de $f'$.

Sustituyendo $w_0 = u_0 - \varepsilon_0$ en la ecuación (6), obtenemos:
$$M'(u_0) = 2rd_1' u_0 \equiv E' \bmod p$$

donde $E' \equiv 2r\varepsilon_0 d_1' - d_2' \bmod p$. Si el polinomio $M'(u_0)$ es no constante, es decir, $d_1' \not\equiv 0 \bmod p$ entonces, fijado $E'$ hay un único $u_0$ verificando esa ecuación.

Definimos $V'(\Delta; a, c)$ como el conjunto de los elementos $v \in F_p$ para los que existe alguna combinación $(d_1', d_2', \varepsilon_0) \in Z^* \times Z^3$ que satisface
$$|d_1'| \leq 9\Delta, |d_2'| \leq 21\Delta^3, |\varepsilon_0| \leq \Delta$$
$$2rd_1' v \equiv E \bmod p$$

Como la cantidad de valores que $E'$ puede tomar está en $O(\Delta^3)$, se sigue que $\#V'(\Delta; a, c) = O(\Delta^4)$.

*Etapa 6: Predicción del generador cuando $f_0 \equiv_p 0$.*

Utilizamos un algoritmo determinista y polinomial (ver [9]) para computar un vector corto $f' \in L'$, y este vector debe ser paralelo a $e'$. Podemos reconstruir con sencillez $e' = f'/f_0'$, lo que nosproporciona $\varepsilon_0$ y consecuentemente $u_0$.

Finalmente, definiendo $U(\Delta; a, c) = V(\Delta; a, c) \cup V'(\Delta; a, c)$, se concluye la demostración.

## 4. Conclusiones y Problemas Abiertos

Obviamente, el resultado presentado es no trivial solamente cuando $\Delta = O(p^{1/4})$. De este modo, aumentando el tamaño para valores $\Delta$ es sumamente importante, incluso considerando la posibilidad de disponer de más de dos aproximaciones consecutivas. En este sentido pensamos que dando $k$ aproximaciones de valores consecutivos, podríamos construir una retícula $L$ de dimensión $s$, de tal forma que el vector asociado $e$ tenga una norma menor que la raíz $2k$-esima del volumen de la retícula $vol(L)^{1/2k}$ (ver ecuación (3)). Heurísticamente, tendríamos que el vector corto proporciona la solución. Estamos implementado estos métodos en un programa C++ usando la librería NTL, ver [18].

Aparentemente, la misma técnica se puede aplicar cuando $a$ ó $c$ (o ambos) son desconocidos.

Como ya hemos mencionado, varios resultados sobre la prediciendode generadores no lineales han sido obtenidos en el reciente trabajo [2]. Sin embargo, requieren excluir un cuerto pequeño conjunto de valores $(a, c)$ excepcionales. Creemos que nuestra idea, puede ser adaptada para soslayar esteconjunto. De todas formas, ciertamente, esto requiere un trabajo futuro. No se sabe como predecir generadores no lineales, en el caso de que el módulo $p$ sea también secreto. Debemos decir que el caso de generador lineal un método heurístico ha sido propuesto en el el trabajo [7]. Sin embargo, no parece inmediato (incluso heurísticamente) que puede ser adaptado a generadores no lineales.

## Referencias

[1] M. Ajtai, R. Kumar and D. Sivakumar, "A sieve algorithm for the shortest lattice vector problem", Proc. 33rd ACM Symp. on Theory of Comput. (STOC 2001), Association for Computing Machinery, 2001, 601-610.

[2] S.R. Blackburn, D. Gómez-Pérez, J. Gutiérrez, I. Shparlinski, "Predicting nonlinear pseudorandom number generators", Mathematics of Computation, (en prensa).

[3] J. Boyar, "Inferring sequences produced by pseudorandom number generators", J. ACM, 36 (1989), 129-141.

[4] J. Boyar, "Inferring sequences produces by a linear congruential generator missing low-order bits", J. Cryptology **1** (1989) 177-184.

[5] E. F. Brickell and A. M. Odlyzko, "Cryptanalysis: A survey of recent results", Contemp. Cryptology, IEEE Press, NY, 1992, 501-540.

[6] M. Grötschel, L. Lovász and A. Schrijver, "Geometric algorithms and combinatorial optimization", Springer-Verlag, Berlin, 1993.

[7] A. Joux and J. Stern, "Lattice reduction: A toolbox for the cryptanalyst", J. Cryptology, **11** (1998), 161-185.

[8] R. Kannan, "Algorithmic geometry of numbers", Annual Review of Comp. Sci., **2** (1987), 231-267.

[9] R. Kannan, "Minkowski's convex body theorem and integer programming", Math. Oper. Res., **12** (1987), 415-440.

[10] D. E. Knuth, "Deciphering a linear congruential encryption", IEEE Trans. Inf. Theory **31** (1985), 49-52.

[11] H. Krawczyk, "How to predict congruential generators", J. Algorithms, **13** (1992), 527-545.

[12] J. C. Lagarias, "Pseudorandom number generators in cryptography and number theory", Proc. Symp. In Appl. Math., Amer. Math. Soc., Providence, RI, **42** (1990), 115-143.

[13] A. K. Lenstra, H. W. Lenstra and L. Lovász, "Factoring polynomials with rational coefficients", Mathematische Annalen, **261** (1982), 515-534.

[14] D. Micciancio and S. Goldwasser, Complexity of lattice problems, Kluwer Acad. Publ., 2002.

[15] P. Q. Nguyen and J. Stern, "Lattice reduction in cryptology: An update", in: W. Bosma (Ed), Proc. ANTS-IV, Lect. Notes in Comp. Sci. Vol. 1838, Springer-Verlag, Berlin, 2000, 85-112.

[16] P. Q. Nguyen and J. Stern, "The two faces of lattices in cryptology", in: J.H. Silverman (Ed), Cryptography and Lattices Lect. Notes in Comp. Sci. Vol. 2146, Springer-Verlag, Berlin, 2001, 146-180.

[17] H. Niederreiter and I. E. Shparlinski, "Recent advances in the theory of nonlinear pseudorandom number generators", in: K.-T. Fang, F.J. Hickernell and H. Niederreiter (Eds), Proc. Conf. on Monte Carlo and Quasi-Monte Carlo Methods, 2000, Springer-Verlag, Berlin, 2002, 86-102.

[18] V. Shoup, "Number theory C++ library (NTL)", version 5.3.1, available at http://www.shoup.net/ntl.